\documentclass[aps,pre,letter,twocolumn,superscriptaddress,floatfix,showpacs,amsmath,amssymb]{revtex4-1}
\usepackage{ifpdf}
 \newif\ifpdf
\ifx\pdfoutput\undefined
   \pdffalse
\else
   \pdfoutput=1
   \pdftrue
\fi
\ifpdf
   \usepackage{graphicx}
   \usepackage{epstopdf}
\else
   \usepackage{graphicx}
\fi

\begin{document}


\title{Two stage order-disorder transformation in simple monatomic supercritical fluid: superstable tetrahedral local order}

\author{R.~E.~Ryltsev}
\affiliation{Institute of Metallurgy, Ural Division of Russian Academy of Sciences, 620016 Yekaterinburg, Russia}

\author{N.~M.~Chtchelkatchev}
\affiliation{Moscow Institute of Physics and Technology, 141700 Moscow, Russia}
\affiliation{L.D. Landau Institute for Theoretical Physics, Russian Academy of Sciences, 142432, Moscow Region, Chernogolovka, Russia}
\affiliation{Department of Physics and Astronomy, California State University Northridge, Northridge, CA 91330, USA}

\date{22 May 2013}

\pacs{61.20.Ne, 65.20.De, 36.40.Qv}

\begin{abstract}
The local order units of dense simple liquid are typically three dimensional (close packed) clusters: hcp, fcc and icosahedrons. We show that the fluid demonstrates the superstable tetrahedral local order up to temperatures several orders of magnitude higher than the melting temperature and down to critical density. While the solid-like local order (hcp, fcc) disappears in the fluid at much lower temperatures and far above critical density. We conclude that the supercritical fluid shows the temperature (density) driven two stage ``melting'' of the three dimensional local order. We also find that the structure relaxation times in the supercritical fluid are much larger than ones estimated for weakly interactive gas even far above the melting line.
\end{abstract}

\maketitle
\section{Introduction}

A supercritical fluid is any substance at a temperature and pressure above its critical point, where distinct liquid and gas phases do not exist. In recent years, the increasing attention in the study of this state of matter has appeared due to the development of supercritical technologies~\cite{Kiran}. The microscopic mechanism that distinguishes supercritical fluid, gas, liquid and solid is one of the central issues that puzzle physicists~\cite{Stillinger,Simeoni,Brazhkin_UFN}. The question if ``the supercritical region of a liquid consists of one single state or not'' is now the matter of debates~\cite{Simeoni,Wallace,Chisolm,Sato,Nishikawa,Brazhkin_UFN}.

The computer simulation of local order is the high precision method that allows to detect subtle changes between slightly different states of the particle system~\cite{Spaepen,Jakse,Speck,Guerdane} and uncover effects lately accessible to real experiment~\cite{Klumov}. Recently the local structure physics brought together simulations and experiment in the nanoplasmonic, quantum optics, soft condensed matter and quantum spectroscopy~\cite{WaterNature2010,Kuhne,RamanSpectra,Klumov}.

\begin{figure}[t]
  \centering
  \includegraphics[width=\columnwidth]{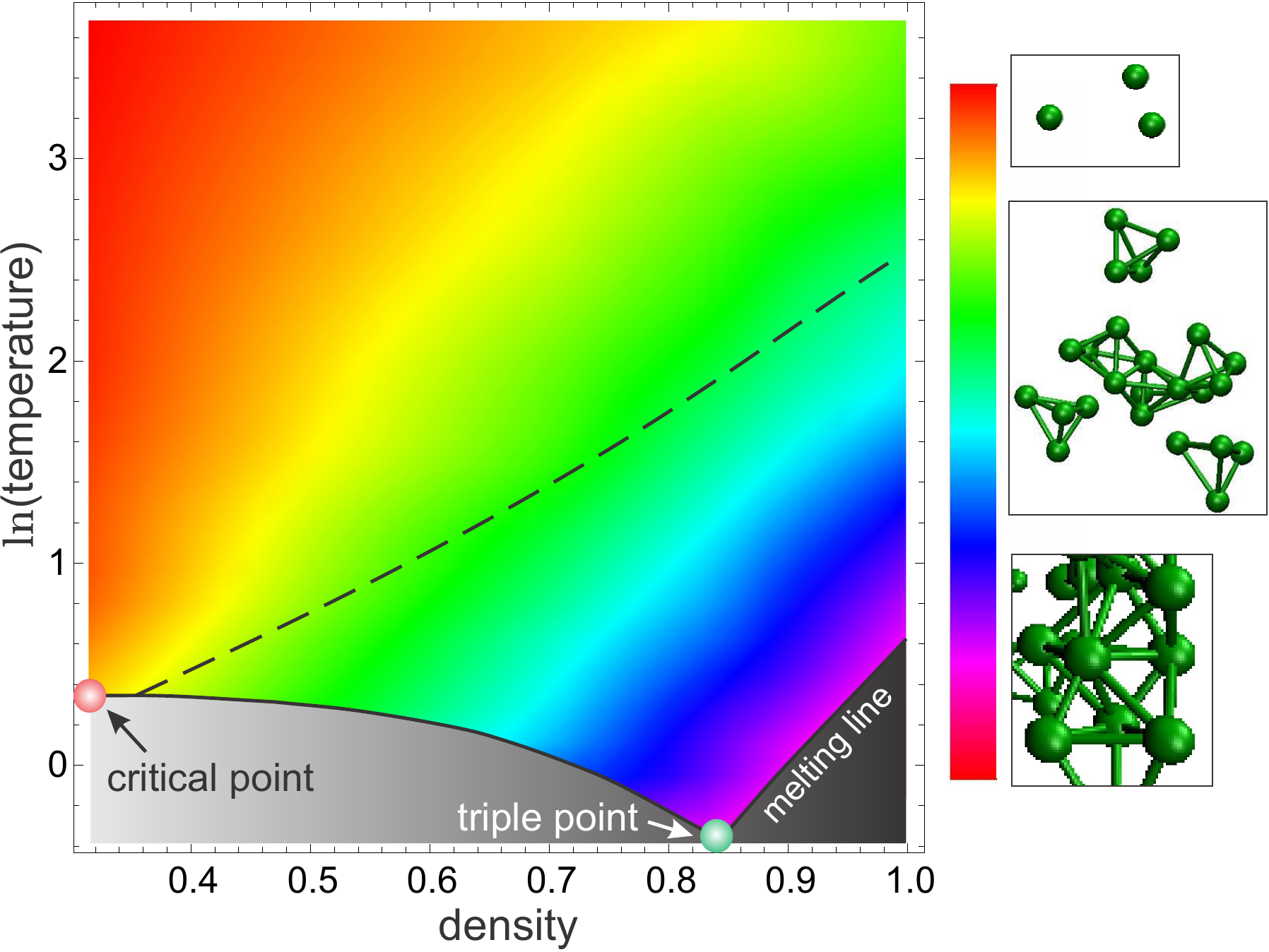}\\
  \caption{(Color online) The density plot of the tetrahedra concentration $P_{\rm tetr}$ in the supercritical simple fluid. The color gradients show $\ln P_{\rm tetr}\in(-3.5,-0.5)$. The snapshots in the insets show the evolution of the ``rigid'' part of the fluid. The dashed line schematically shows the position of the ``dynamical line''~\cite{Brazhkin_UFN}. The probability to find a tetrahedral cluster falls below $0.1$ in the yellow zone (hcp-clusters melt in blue-zone). The grey zone corresponds to the state other than fluid. }\label{Fig1}
\end{figure}

\begin{figure}[h]
  \centering
  \includegraphics[width=0.7\columnwidth]{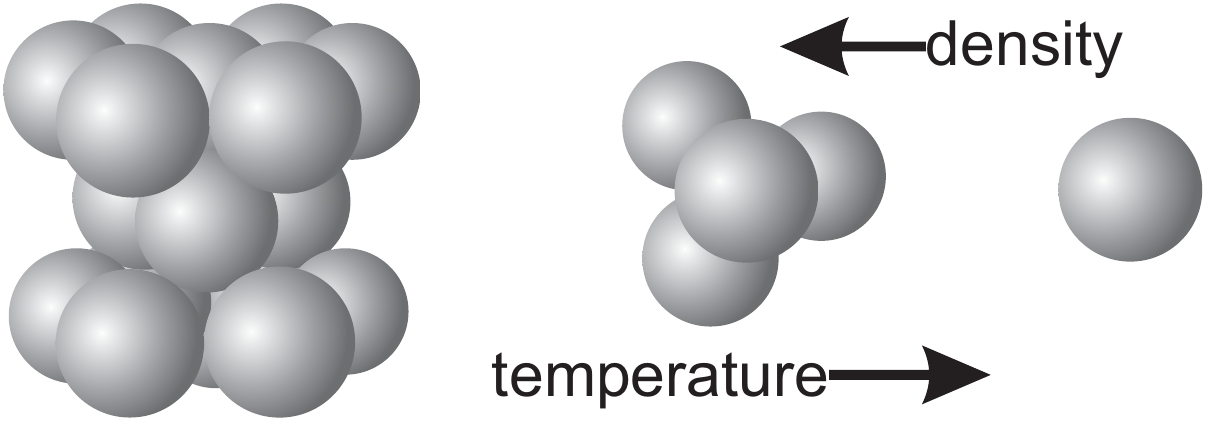}
  \caption{Qualitative evolution of the \textit{three dimensional} local symmetry of simple monatomic supercritical fluid when temperature increases or (and) density decreases (starting from the liquid state near melting line).}\label{Fig0}
\end{figure}

The local order units of dense simple liquid are typically three dimensional (close packed) clusters (hcp, fcc and icosahedrons)~\cite{Patashinskii0,Patashinskii1,Patashinskii2,Nelson_inv,Medvedev}. It is intuitively clear that these clusters, inherent to the liquid state, disappear in the fluid at high temperatures and low densities, see e.g., Ref.~\cite{Patashinskii2} and refs. therein. We confirm that and find out that solid-like local order dominates in the simple monatomic dense liquid (hcp, in the Lennard-Jonnes (LJ) liquid) near the melting line, but disappears in supercritical region at temperatures several times higher the melting temperature or (and) at densities less than about $2\rho_c$, where $\rho_c$ -- critical density. The unexpected issue we see is that small three-dimensional clusters with higher symmetry survive in fluid at extremely high temperatures and low densities: We see tetrahedra in simple monatomic supercritical fluid. They survive up to the temperatures several orders of magnitude higher than the melting temperature and down to $\rho_c$ (see Fig.~\ref{Fig1}). So the supercritical fluid shows the temperature (density) driven two stage ``melting'' of the three dimensional local order (see Fig.~\ref{Fig0}).

Qualitatively it can accounted for as follows: at high enough densities, in the LJ-liquid and in the fluid near the melting line the number of the nearest neighbours is $12\pm1$. That favours the local clusters with the closed-packed structure: hcp for LJ-system~\cite{Patashinskii0,Patashinskii1}. Temperature increase or (and) density decrease causes the reduction of mean nearest neighbour number (see Fig.~\ref{fig:fractions}c) and the growth of its fluctuations (see insert in Fig.~\ref{fig:fractions}c) that destroys the closed-packed local order but still allows the tetrahedral one.

The local cluster in the fluid already changes its symmetry when at least one of its particles moves at the distance of the order of $r_0\delta$, where $\delta\lesssim 0.1\sim 1/\bar n_b$, $\bar n_b$ is the mean number of the nearest neighbours and $r_0$ is the mean distance between the nearest neighbours. The correlations between particles in the fluid are rather weak (compared to the liquid) and so one can naively estimate the life time of the local cluster as $\tau_{\rm sct} \delta$, where $\tau_{\rm sct}$ is the particle mean scattering time. We show that the local structure in the supercritical fluid, even at temperatures several orders of magnitude higher than the melting temperature, survives at the time scales by the order of magnitude higher than this naive estimate. These data we extract from the time pair correlation function of the local structure (bond orientational) order parameters.

A supercritical fluid is typically characterized as the state ``in between'' a gas and a liquid. The characteristic property of a liquid is the existence of the local structure and relative long structural relaxation time scales (typically larger than $\tau_{\rm sct}$). It follows that the supercritical fluid preserves the local structure up to very high temperatures. However this local structure shows itself only at relatively high frequency scales (compared to that in a liquid) $\omega\lesssim 1/\tau_{\rm sct}$. At temperatures where the fraction of tetrahedra becomes less than 1$\%$, the tetrahedra life time starts approaching towards the naive weakly interactive gas estimate, $\tau_{\rm sct} \delta$. This is natural since in this temperature region the supercritical fluid is already ``more gas than liquid''.

\begin{figure}[b]
  \centering
  \includegraphics[width=\columnwidth]{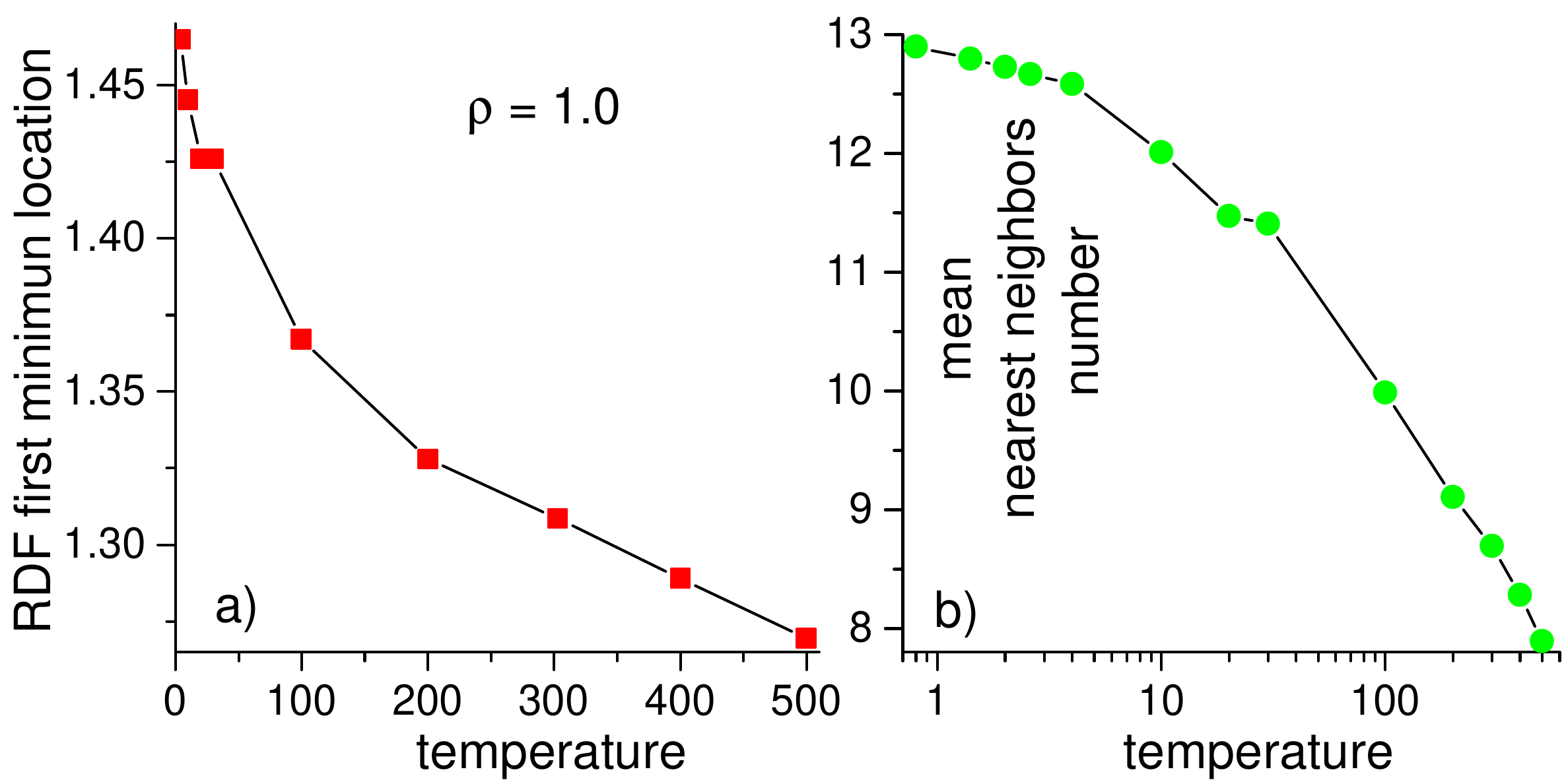}\\
  \caption{(Color online) (a) The temperature dependence of the first RDF minimum location at density $\rho=1$. (b) the mean nearest neighbour number determined by the cut-off radius obtained from the first RDF minimum.}\label{figRDFminimum}
\end{figure}
Recently the conjecture has been made in Ref.~\cite{Brazhkin_UFN} that liquids in the the supercritical region may exist in two qualitatively distinct dynamical states, different by some fitches of the molecule motion at small time scales. It would be natural to expect that the transformation of the local structures of the fluid follows the dynamical crossover line \cite{note}. But we see different situation, as follows from Fig.~\ref{Fig1}.

\section{Simulation procedure}

For molecular dynamic (MD) simulations, we have used $\rm{DL\_POLY}$  Molecular Simulation Package~\cite{dlpoly} developed at Daresbury Laboratory.

 For simulations we mostly used the LJ pair potential model in a wide range of parameters and checked the stability of the main results going to the Soft Spheres model. For LJ liquid we apply the standard pair potential~\cite{Lennard-Jones,Hansen-McDonald}, $U(r)=4\varepsilon [(\sigma/r)^{12}-(\sigma/r)^{6}]$, where $\varepsilon$ – is the unit of energy, and $\sigma$ is the core diameter. In the remainder of this paper we use the dimensionless quantities: $\tilde r= r/\sigma$, $\tilde U = U/\varepsilon$, temperature $\tilde T = T/\varepsilon$, density $\tilde{\rho}\equiv N \sigma^{3}/V$, and time $\tilde t=t/[\sigma\sqrt{m/\varepsilon}]$, where $m$ and $V$ are the molecular mass and system volume correspondingly. As we will only use these reduced variables, we omit the tildes. For the soft sphere model~\cite{Heyes} we apply $U(r)=4\varepsilon (\sigma/r)^{12}$.

For simulations, we have considered the system of $N=30000$ particles that were simulated under periodic boundary conditions mostly in the Nose-Hover (NVT) and also in NPT ensambles. The MD time step was $t=0.001$ that provides good energy conservation. The system was studied in the density region of $\rho\in(0.32-1.0)$ at temperatures up to T = 100 (that is much higher than the melting line and the critical temperature). According to equilibrium temperature-density phase diagram~\cite{BSmit,Agrawal,Kalyuzhnyi,Goddard,Mastny,Akmed,Khrapak,Morfill,Asano,Watanabe,Frenkelbook}, this range completely includes the area corresponding to liquid phase and widely covers the region of fluid state.

\section{Local structure.}

\subsection{The nearest neighbour list and SANN algorithm}

The key problem when studying local order is the determination of the nearest neighbors list for each particle. The generally accepted way to do it is the fixed-distance cutoff method when two particles, located at points ${\bf r_1}$ and ${\bf r_2}$, are considered as neighbors if $\left| {{\bf r}_2  - {\bf r}_1 } \right| < r_{{\rm cut}}$, where $r_{{\rm cut}}$ -- cutoff radius. The main problem on this way is the $r_{\rm cut}$ evaluation. The widespread method is equating cutoff radius with the location of first minimum of radial distribution function (RDF) ~\cite{Frenkelbook} that gives temperature dependent $r_{\rm cut}$ at fixed density (Fig.~\ref{figRDFminimum}a). Unfortunately this method reveals unphysical temperature dependencies of different structural characteristics, such as average nearest neighbors numbers, see Fig.~\ref{figRDFminimum}b. Much better results gives the choice of the temperature independent cutoff radius at given density. However the method of $r_{\rm cut}(\rho)$ evaluation is still unclear~\cite{Nelson_inv}. For LJ system at high enough density (more that triple point density 0.84), reasonable results can de obtained by equating $r_{\rm cut}$ with the location of first RDF minimum for corresponding low temperature crystal state. But the way of using this trick in the low density region is unclear since there is no stable homogeneous crystal phase there.

\begin{figure}[t]
  \centering
  \includegraphics[width=\columnwidth]{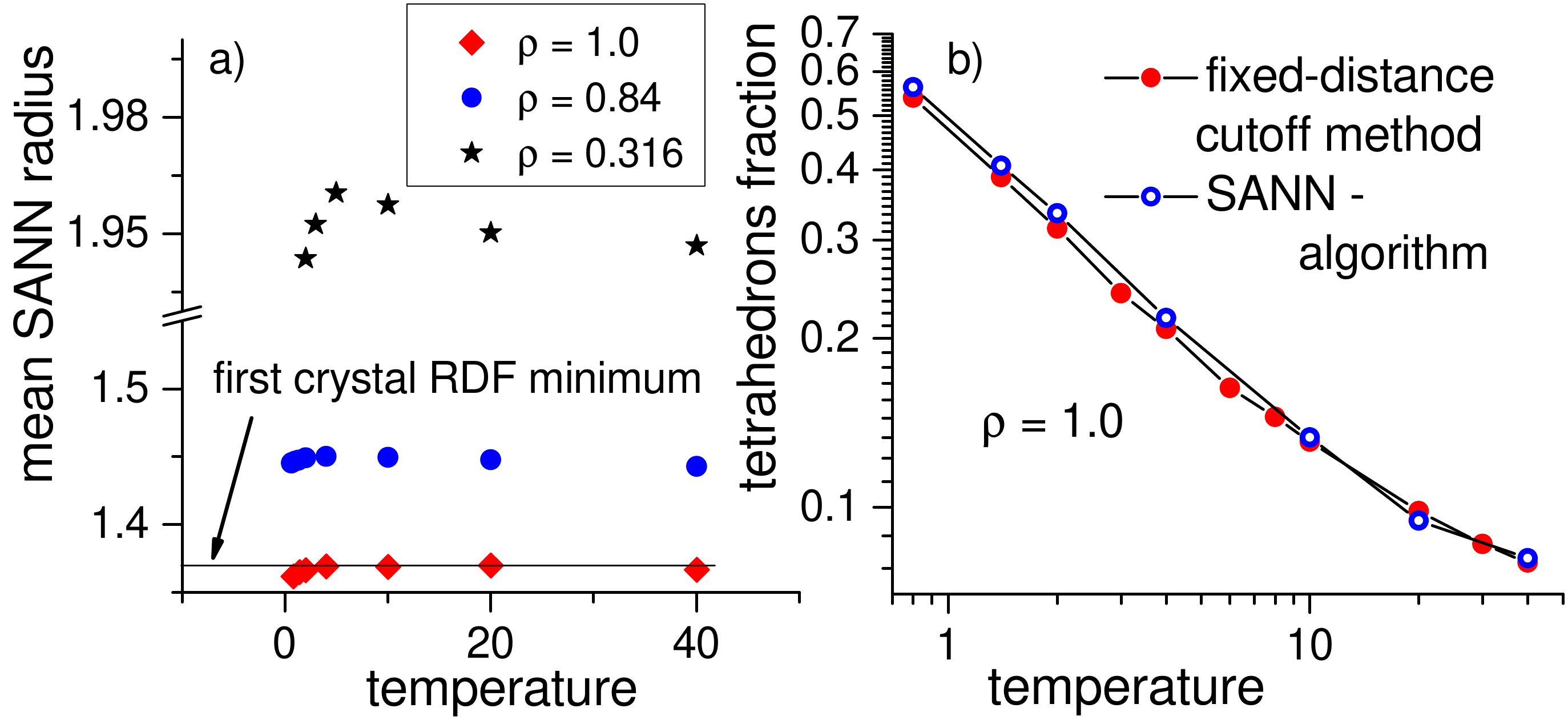}
  \caption{(Color online) (a) The temperature dependence of the nearest neighbour cut-off radius $r_{\rm cut}$ obtained using the SANN algorithm for two choices of the densities. It follows that $r_{\rm cut}$ slightly depends on $T$. (b) The comparison of the temperature dependencies of fraction of the tetrahedral clusters obtained using the fixed distance cut-off method and the SANN algorithm correspondingly.} \label{figSANN}
\end{figure}

Recently, a universal parameter-free algorithm was proposed to determine the neighbor list for particle-systems~\cite{SANN}. This solid-angle based nearest-neighbor algorithm (SANN) attributes to each possible neighbor a solid angle and determines the cutoff radius by the requirement that the sum of the solid angles is $4\pi$~\cite{SANN}. We check that, this algorithm gives weak temperature dependence of cutoff radius for dense LJ liquid as well as for low-dense fluid (see Fig.~\ref{figSANN}a). Moreover, for dense liquid, mean SANN radius is practically the same as the location of first RDF minimum for corresponding crystal state (Fig.~\ref{figSANN}a).  The results of local order analysis also coincide for both methods (Fig.~\ref{figSANN}b). It suggests that SANN is valid for systems with arbitrary density and so all our results concerning the local structure were obtained using this algorithm.

\subsection{Cluster symmetry determination}
With the nearest neighbor lists one can perform the local order analysis. The main characteristics we are interested in are the fractions of atoms, possessing the certain type of local structure, and their spatial distributions. As the candidates for the structural units we choose close packed clusters (hcp, fcc and icosahedron) and the tetrahedron as the most compact (and stable) unit for system with isotropic potential.

As the basic tool for determination of local structure we use bond orientational order parameters $q_l({\bf r})$ defined as~\cite{Patashinskii0,Patashinskii1,Nelson_inv}:
\begin{equation}\label{invariants_def}
q_l^2 ({\bf r}) = \frac{{4\pi }}{{2l + 1}}\sum\limits_{m =  - l}^l {\left| {\frac{1}{{n_b ({\bf r})}}\sum\limits_{\rm bonds} {Y_{lm} (\theta,\varphi)} } \right|} ^2.
\end{equation}
Here $(\theta,\varphi)$ -- are respectively polar and azimuthal angles of the nearest neighbors radius-vectors for the particle located at point ${\bf r}$; $n_b ({\bf r})$ is the number of nearest neighbors for this particle; $Y_{lm} (\theta, \varphi)$ are spherical harmonics. The sign $\sum\nolimits_{\rm bonds}$ means the sum over nearest neighbors of given particle. The values of order parameters (\ref{invariants_def}) are explicitly determined for any ideal geometrical figure and do not depend on spatial cluster orientation that allows using them as local structure indicators. Among the parameters (\ref{invariants_def}), the $q_6$  is typically one of the most informative one so we use it basically in this work.

 The criterions we use for define the atom ${\bf r}$ and their nearest neighbors as close packed cluster are:
\begin{gather}\label{eqclosepacked}
{\left| {q_l ({\bf r})/{{q_l^{\rm (id)} }}  - 1 } \right|}\le \delta, \quad n_b({\bf r})=12,
\end{gather}
where $ q_l^{\rm(id)}$ is the bond order parameter value for corresponding ideal cluster. Similarly, we treat each group of four atoms [triangle pyramid with the vertex at point ${\bf r}$] as tetrahedron if they are nearest neighbors of each others and each of vertex angle $\phi_k ({\bf r})$ ($k=1,2,3$) satisfies the criterion:
\begin{equation}\label{tetrahedron criteria}
\left| {\phi _k ({\bf r})/\phi ^{\rm(id)}  - 1} \right| < \delta,
\end{equation}
where $\phi^{\rm(id)}=\pi/3$ is vertex angle for regular tetrahedron.

\begin{figure}[b]
  \centering
   \includegraphics[width=\columnwidth]{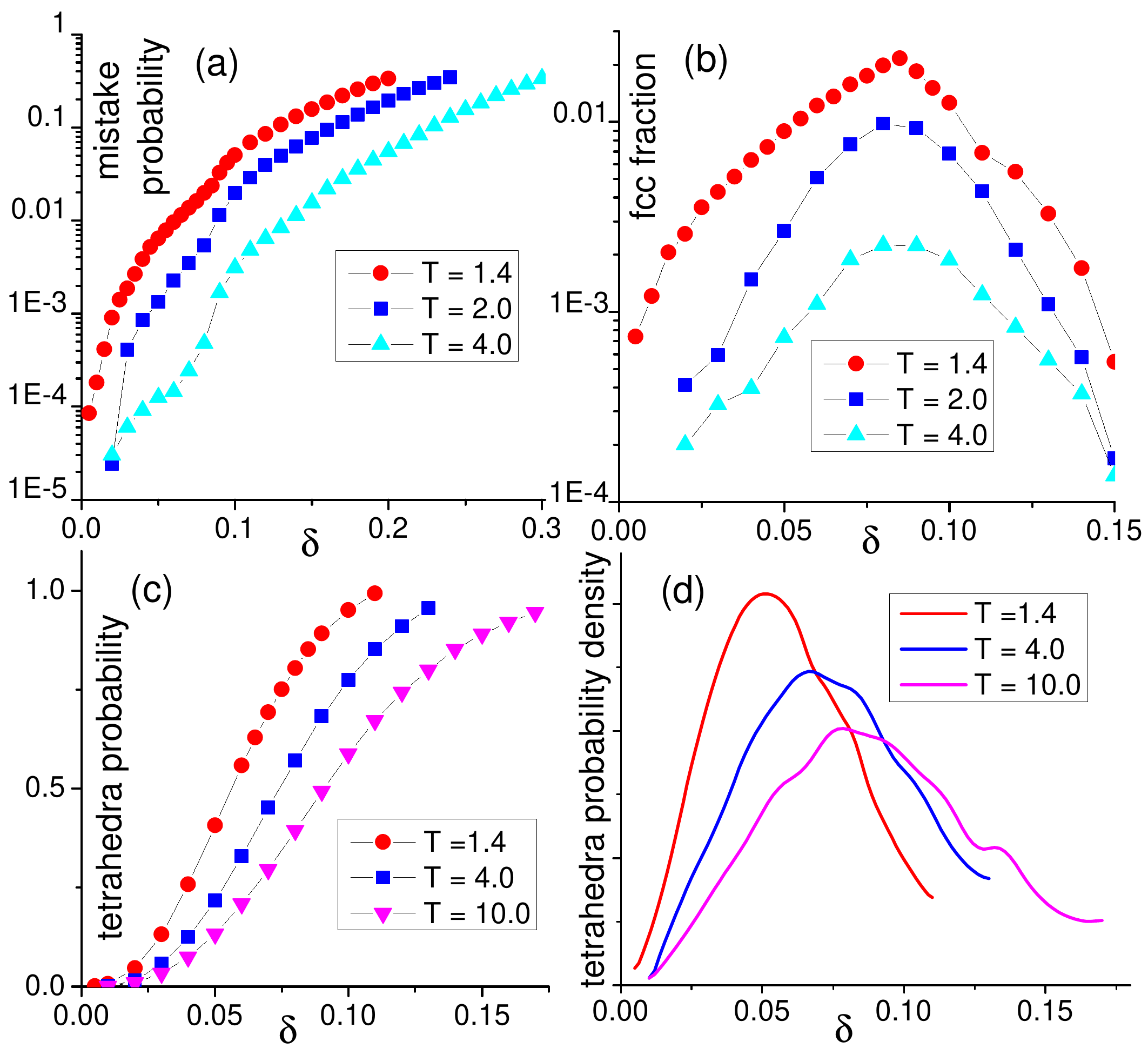}
  \caption{(Color online) (a) The dependencies of probability to make error in local structure recognition on $\delta$; (b) The temperature dependencies of the fractions of fcc-ordered atoms on $\delta$. The probability functions: (c) and  the probability densities (d) for tetrahedral clusters distorted with the parameter $\delta$. It follows that the probability has an extremum at certain $\delta$. Our choice of $\delta$ when we distinguish perfect and imperfect clusters corresponds to the maximum of the probability density at temperature corresponding to the melting line. It should be noted that the probability densities for tetrahedral clusters are expected to be Gaussian-shaped and so the additional maxima and bends we see are probably the numeric differentiation effects.}\label{fig_delta_hcp}
\end{figure}
Let us discuss the criteria we have used to distinguish the nearest neighbour clusters (the choice of $\delta$). For perfect clusters each symmetry exactly corresponds to particular value of $q_l$. If the clusters are slightly disordered (imperfect) then one can plot the histogram for $q_l$ and determine the symmetry by the maxima of the histogram. However with the increase of the disorder $q_l$-histogram fail to distinguish the nearest neighbour symmetry because of the ``symmetry overlap problem''~\cite{Patashinskii1}. If the imperfection of the clusters are larger then some critical value then $q_l$-criteria cannot well distinguish the cluster symmetry so choosing $\delta$ we kept in mind this critical disorder. Fig.~\ref{fig_delta_hcp}a shows the dependencies of the probability to make error in local structure recognition on $\delta$. We see that there is a kink at $\delta\approx 0.85$ corresponding to drastic overlap increasing. In Fig.~\ref{fig_delta_hcp}b we see the consequence of this overlapping: the fraction of fcc-ordered atoms starts to decrease at this value of $\delta$.  So we choose $\delta=0.05$ slightly below this threshold.

To fix the tetrahedrons we have used similar ideas. It is impossible to tile the 3D space by perfect tetrahedra however the ensemble of distorted enough tetrahedra can tessellate the 3D space. Fig.~\ref{fig_delta_hcp}(c) and (d) shows the probability functions  and  the probability densities for tetrahedral clusters distorted with the distortion parameter $\delta$. It follows that the probability has extremum at certain $\delta$ corresponding to its mean value. We see that the dispersions of these distributions increase with temperature that makes the mean value of $\delta$ to be ill-defined at temperatures much higher than the melting point. So our choice of $\delta$, when we distinguish perfect and imperfect tetrahedra, approximately corresponds to the maximum of the probability density at temperature corresponding to the melting line.

\subsection{Local order analysis}

\begin{figure}
  \centering
  \includegraphics[width=0.99\columnwidth]{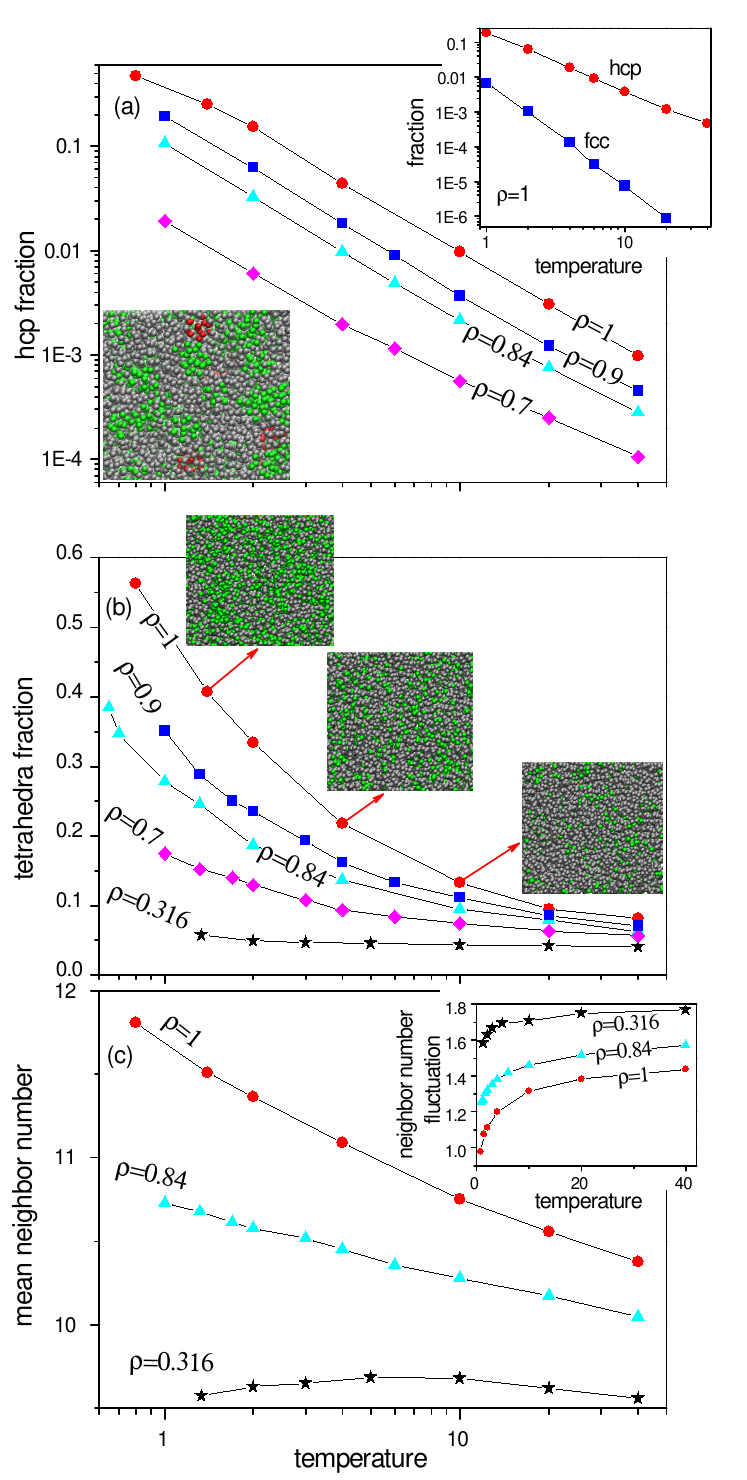} 
  \caption{(Color online) (a) Main frame: the temperature dependencies of hcp-ordered atoms fractions at different densities. Upper insert: the comparison of temperature dependencies of hcp-ordered (bullets) and fcc-ordered (squares) atoms fractions at $\rho=1.0$. Lower insert: the snapshots of atom structure at $\rho=1.0$, $T=1.4$. The hcp-ordered atoms are green, the fcc-ordered ones are red and the remainder are gray.  (b) Main frame: the temperature dependencies of tetrahedrally-ordered atoms fractions at different densities. Snapshot inserts show the atom structure at $\rho=1.0$ and temperatures indicated by arrows. The atoms with tetrahedral order are colored green and the remainder are gray. (c) The temperature dependencies of mean nearest neighbor numbers (main frame) and their fluctuations (inset) at different densities. For all graphs, (red) bullets -- $\rho=1$, (blue) squares -- 0.9, (cyan) triangles -- 0.84 (triple point), (magenta) diamonds -- 0.7, (black) stars -- 0.316 (critical point).}\label{fig:fractions}
\end{figure}

Using Eq.~\eqref{invariants_def} we calculated mean fractions of atoms with different types of local structure in wide ranges of temperatures and densities (see Fig.~\ref{fig:fractions}). We see that close packed local order is mainly presented by hcp-clusters, the fraction of fcc-clusters is less on the order of magnitude (see insert in Fig.~\ref{fig:fractions}a) and there are no icosahedra at all. The solid-like local order (hcp, fcc) is pronounced only for dense liquid and disappears in the fluid at temperatures of the order of the 2-3 the melting temperature (Fig.~\ref{fig:fractions}a) or (and) densities  $\rho \lesssim 2\rho_c$. On the other hand, the tetrahedral order is much more stable and survives up to the temperatures several orders of magnitude higher than the melting temperature and down to $\rho_c$ (Fig.~\ref{fig:fractions}b).

In Fig.~\ref{fig:fractions}c we show that in the dense LJ-liquid (and in the dense fluid near the melting line) the number of the nearest neighbours $n_b$ is $12\pm1$. That favours the local clusters with the closed-packed structure (hcp for LJ-system)~\cite{Patashinskii0,Patashinskii1}. In Fig.~\ref{fig:fractions}c we show that the temperature increase or (and) density decrease leads to the reduction of $n_b$ and the growth of its fluctuations (see insert in Fig.~\ref{fig:fractions}c). The local order with large $n_b$ is more sensitive to $n_b$ fluctuations.  So tetrahedron appears to be one of the most stable and tight element of the local structure and so it effectively plays the role of the structural ``quantum'' of the fluid.
\begin{figure}[t]
  \centering
  \includegraphics[width=0.95\columnwidth]{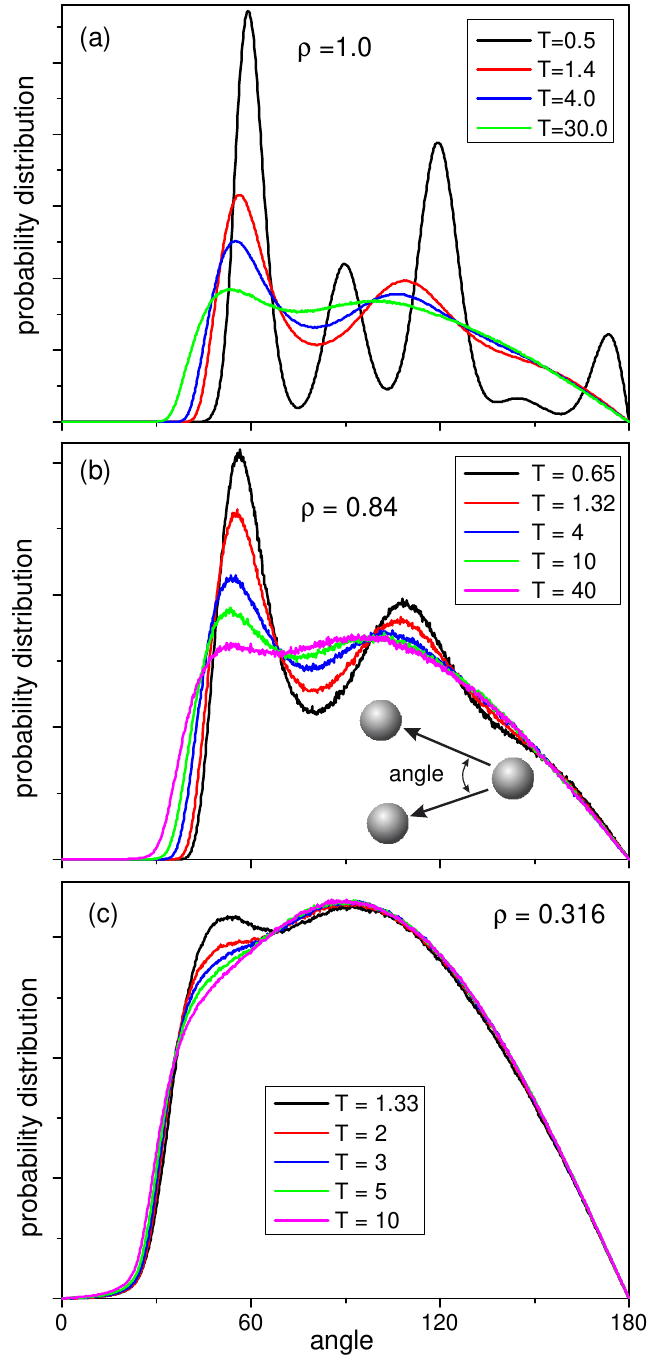}\\
    \caption{(Color online) Angular distribution for Lennard-Jones system at $\rho=\{1.0,0.84,0.316\}$ and different temperatures. The insert in (b) explains the meaning of the the angle.} \label{fig:angle}
\end{figure}
\begin{figure}[t]
  \centering
  \includegraphics[width=0.67\columnwidth]{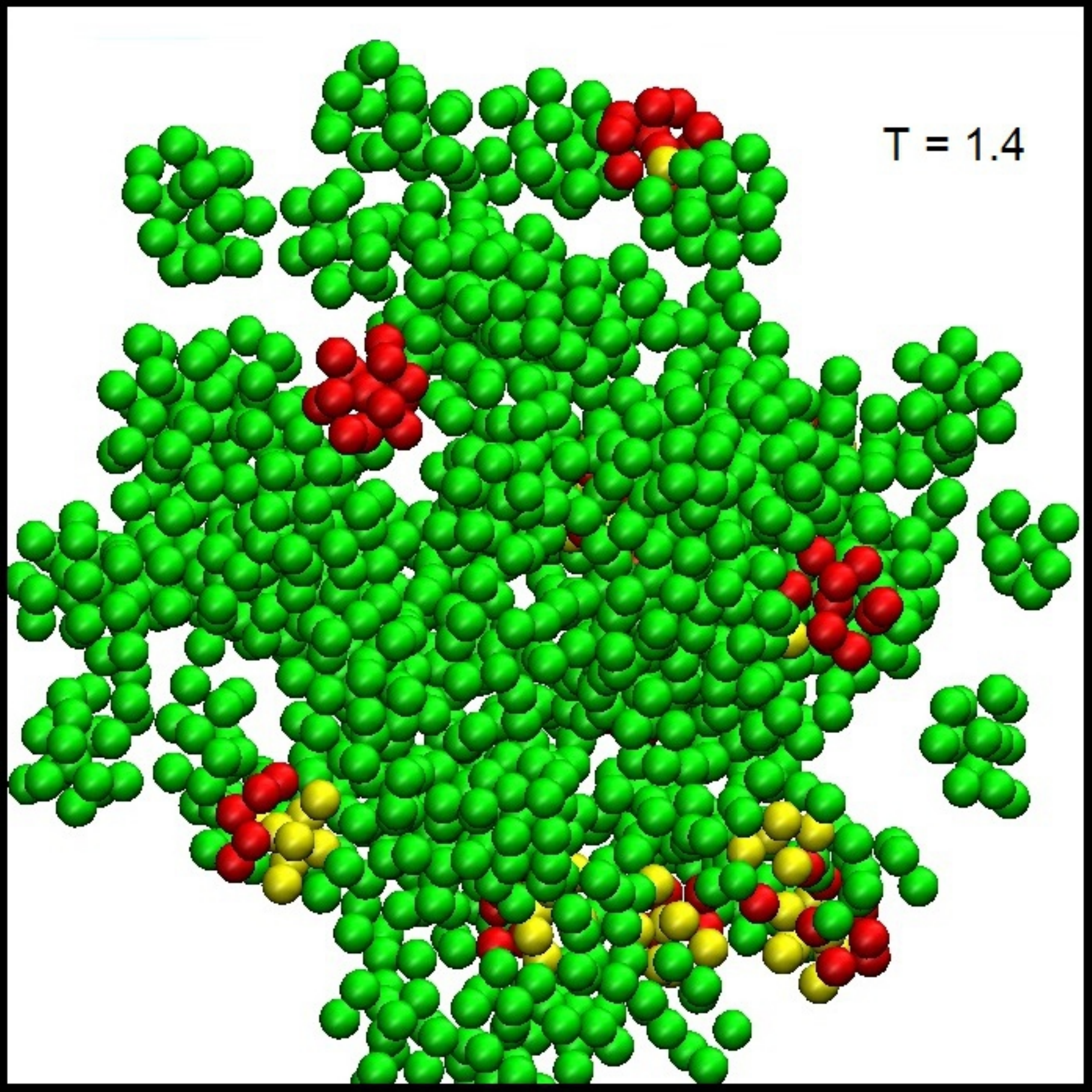}\\
   \includegraphics[width=0.67\columnwidth]{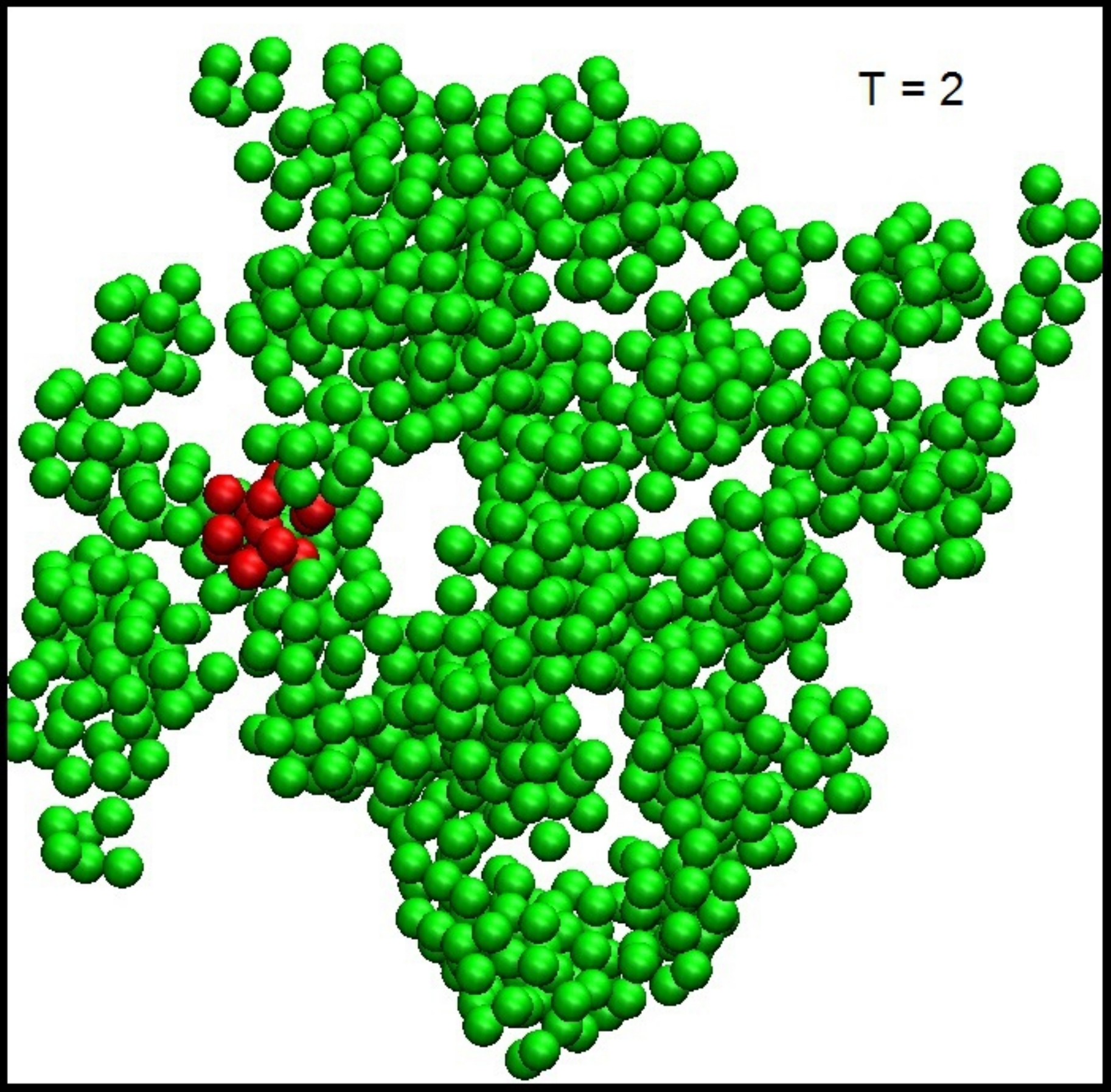} \\ \includegraphics[width=0.67\columnwidth]{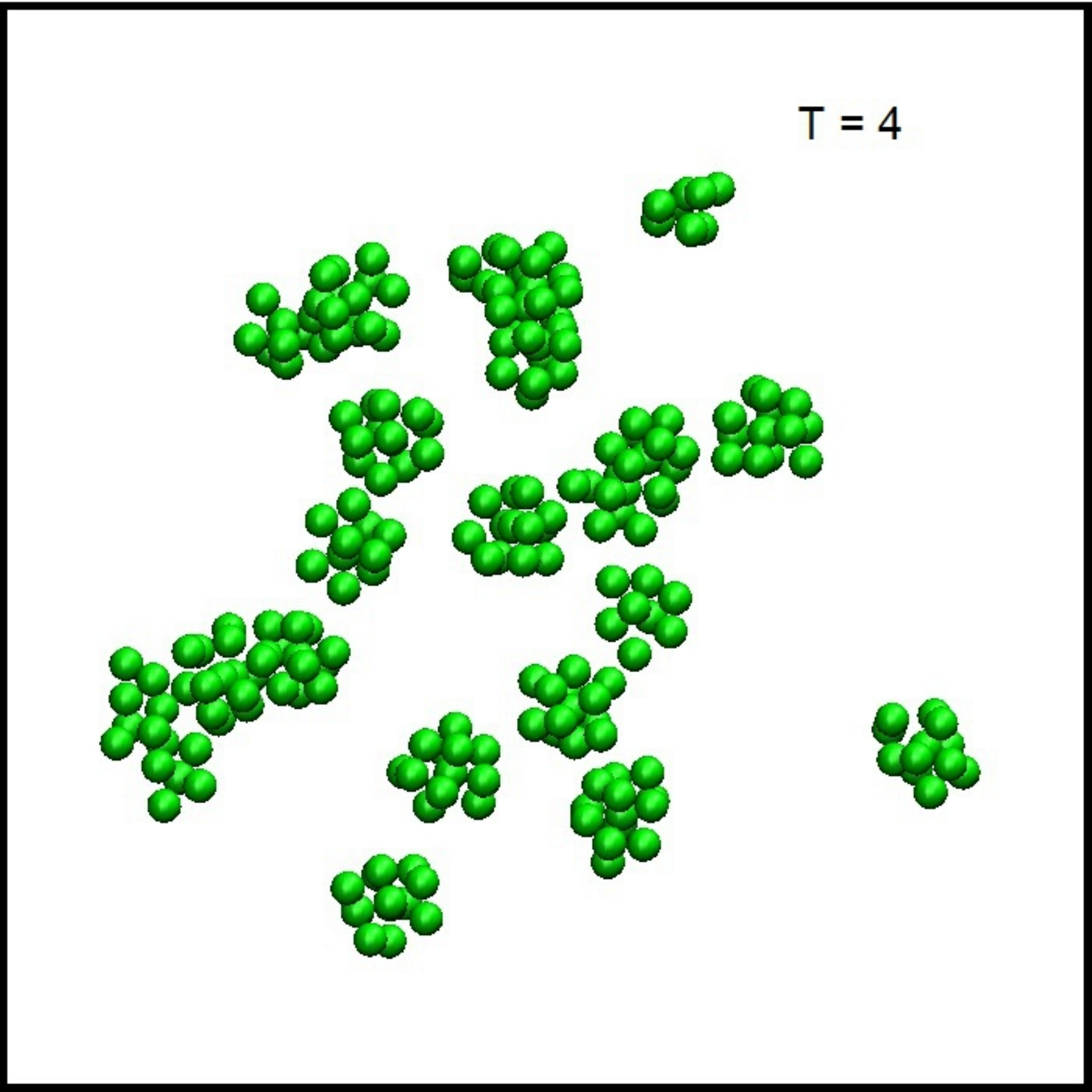}
  \caption{(Color online) Snapshots of the local ordered particles in LJ liquid at the density $\rho=1$ and the temperatures: $T=(1.4; 2.0; 4.0)$. Green color denotes hcp clusters, red --- fcc, and yellow --- the other locally ordered atoms that may belong to both structures (the bond order parameters slightly overlap for these atoms). The atoms with the structureless local order are completely removed.}\label{fig:hcp_snapshot}
\end{figure}
Qualitatively, the tetrahedral local order can be presented in fluid regardless of any parameters like $\delta$: by angular distribution function of fluid particles. Such distribution for Lennard-Jones system was firstly calculated thirty yeas ago by Belashenko~\cite{Belashenko}. We reproduced results of Belashenko in order to make more accurate and smooth curves (see Fig.~\ref{fig:angle}). We see that, at high densities, there is a clear maximum near perfect tetrahedral angle $\pi/3$ that demonstrates strong tendency to tetrahedral formation. The distribution at the density $\rho=0.316$ does not have pronounced feature at 60$^\circ$. However the absence of the pronounced maxima in the angle distribution does not automatically mean the absence of tetrahedra. We emphasize that the concentration of tetrahedra is rather small at this density. However the fluctuation of the concentration is still much smaller than the concentration.

\subsection{The spatial distribution of local order clusters}

Another interesting issue is the spatial distribution of the clusters with different types of local order. To study these clusters we colour atoms in accordance with the local ordering. For close packed clusters the atom $\bf r$ and its nearest neighbours are colored if the value of $q_6({\bf r})$ satisfies the criteria (\ref{eqclosepacked}). It is clear that some atoms can simultaneously belong to different types of local order and so we called them ''contested'' atoms and paint special colour. Fig.~\ref{fig:hcp_snapshot} shows the configuration shapshots with atoms coloured this way. Green color denotes hcp clusters, red --- fcc, and yellow --- the contested ones. The atoms with the structureless local order are completely removed for visual clarity. We see that the structurally ordered atoms are not uniformly distributed but rather tend to associate in branched ''superclusters'' with (may be) fractal structure. The detailed investigation of this issue is the matter of separate work.

\begin{figure}[t]
  \centering
  \includegraphics[width=0.99\columnwidth]{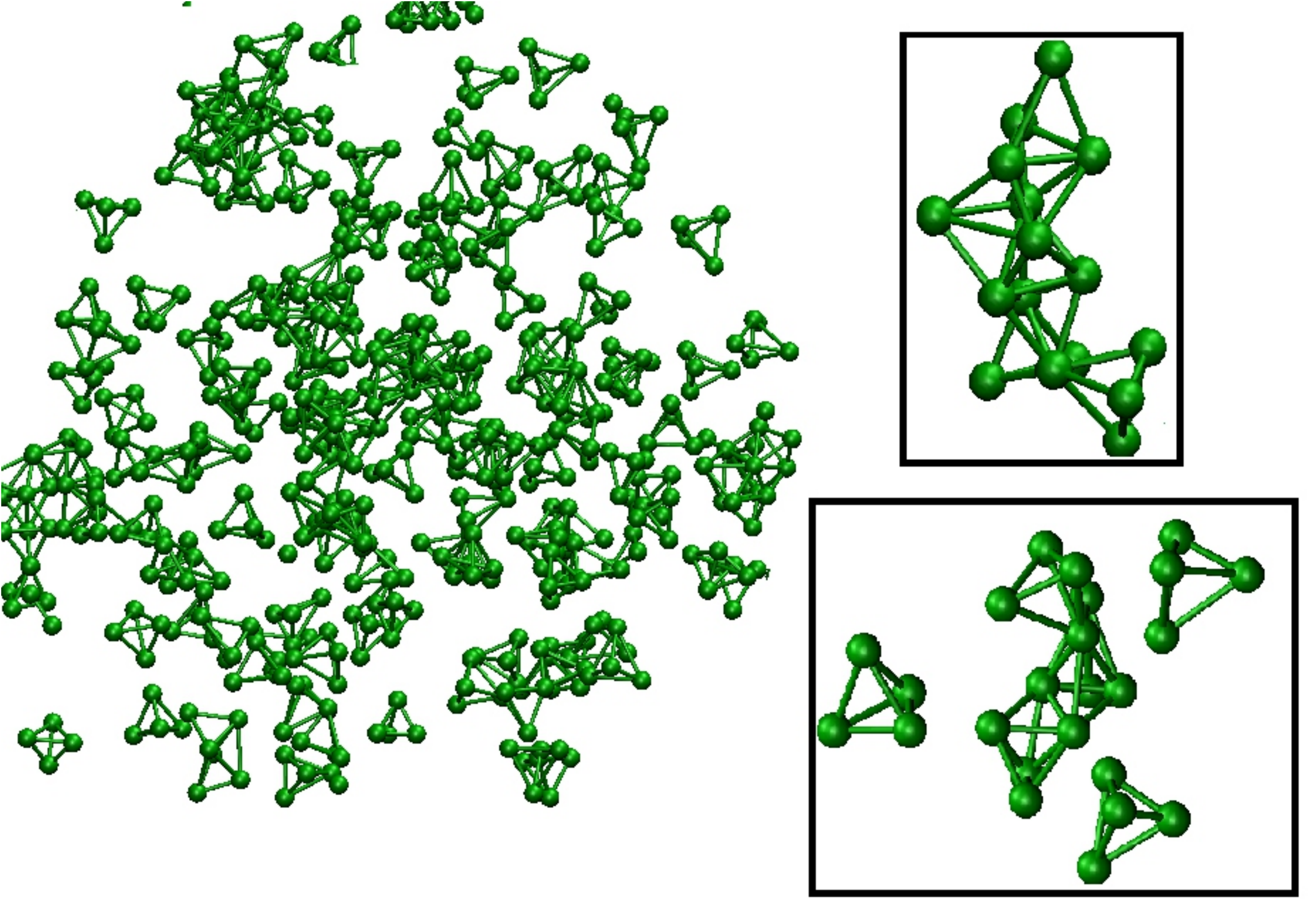}
  \caption{(Color online)Snapshots of the tetrahedrally ordered particles in LJ liquid at the density $\rho=1$ and the temperatures  $T=10$. The atoms with the non-tetrahedral local order are completely removed.}\label{fig:tet_snapshot}
\end{figure}

For tetrahedra visualization we use similar ideas. In this case we have the only local structure element -- tetrahedron, and so there is no need in special coloring. So we just delete the atoms with non-tetrahedral order and connect the remainder ones by bonds (Fig.~\ref{fig:tet_snapshot}). For visual clarity we show the snapshots corresponding to high enough temperature $T=10$ at which the tetrahedra fraction relatively small. We see that, despite of high temperature, the formation of the polytetrahedral clusters takes place (see inserts in Fig.~\ref{fig:tet_snapshot}).

\section{Characteristic time scales.} Another important issue is the local order lifetime. We extract it from the autocorrelation function of the local structure:

\begin{figure}[b]
  \centering
  \includegraphics[width=\columnwidth]{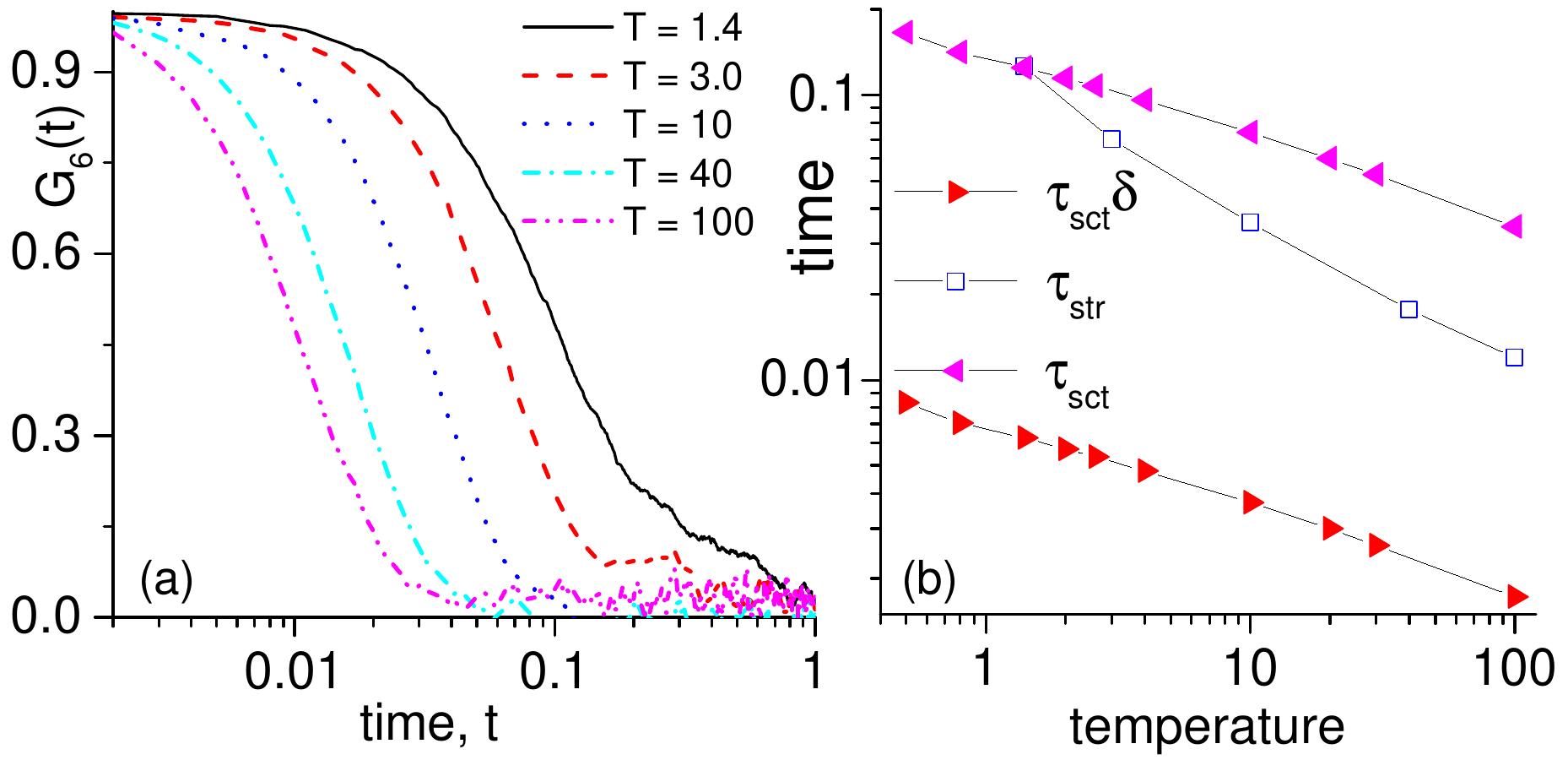} 
  \caption{(Color online) (a) The order parameter correlation function $G_6(t)$ at temperatures $T=1.4-100$. The characteristic decay time we identify with $\tau_{\rm str}$.  (b) Temperature evolution of the local order life time,  $\tau_{\rm str}$. At the same graph we show the effective Einstein vibration period that we relate with  $\tau_{\rm sct}$. At the melting line  $\tau_{\rm sct}\approx  \tau_{\rm str}$. To investigate the difference between these time scales at larger temperatures we normalized $\tau_{\rm sct}$ making the graph, so $\tau_{\rm sct}= \tau_{\rm str}$ exactly at the melting line. The lower graph shows $\tau_{\rm sct}\delta$ -- the estimate of the cluster lifetimes in the supercritical fluid obtained within the weakly interactive gas model. }\label{figGl}
\end{figure}

\begin{equation}\label{<Q(t)Q(0)>}
G_{l} (t) = \frac{\langle q_l (t+\tau,\textbf{r})q_l (\tau,\textbf{r})\rangle_{\tau,\textbf{r}}-\langle q_l(\tau,\textbf{r})\rangle_{\tau,\textbf{r}}^2}{\langle q_l^2(\tau,\textbf{r})\rangle_{\tau,\textbf{r}}-\langle q_l(\tau,\textbf{r})\rangle_{\tau,\textbf{r}}^2 },
\end{equation}
where $\langle\ldots\rangle_{\tau,\textbf{r}}$ is the average over coordinates and time. Using the relation $G_l (\tau _{{\rm str}} ) = 1/e$ we obtain the structure relaxation time $\tau_{\rm str}$ that determines the desired local order life time.

The graphs of $G_6(t)$ at $\rho=1$ and different temperatures are shown in Fig.~\ref{figGl}a. The decay of $G_6(t)$ is almost exponential at long times and relaxation time $\tau_{\rm str}$ demonstrates predictable decrease with temperature (Fig.~\ref{figGl}b). We see that the local structure correlations survive at time scales that appear to be of the order of scattering time $\tau_{\rm sct}$ (Fig.~\ref{figGl}b). The value of $\tau_{\rm sct}$ we extract from the estimation $\tau_{\rm sct}\simeq T_{\rm vib}/4$, where $T_{\rm vib}=2\pi/{\Omega_0}$ is the period at which the tagged particle would vibrate in the ''cage'' of its nearest neighbors \cite{Hansen-McDonald}. The corresponding frequency $\Omega_0$, that is effective Einstein vibration frequency, can be obtained for system with pair potential as~\cite{Hansen-McDonald}:
\begin{equation}\label{Omega}
\Omega _0^2  = \frac{\rho }{{3m}}\int {\nabla^2 U(r)} g(r)d{\bf r},
\end{equation}
where $g(r)$ is the radial distribution function.

The effective Einstein vibration period that we relate with $\tau_{\rm sct}$ is expected to be  $\tau_{\rm sct}\approx\tau_{\rm str}$ at the melting line. To investigate the difference between these time scales at larger temperatures we normalized $\tau_{\rm sct}$ making the graph, so $\tau_{\rm sct}= \tau_{\rm sct}$ exactly at the melting line (Fig.~\ref{figGl}b). The lower curve in this graph shows $\tau_{\rm sct}\delta$ -- the estimate of the cluster lifetimes in the supercritical fluid obtained within the weakly interactive gas model, where $\delta\lesssim 0.1\sim 1/\bar n_b$. It follows that at temperatures larger than $100$ melting temperatures the structural life time develops with temperatures towards $\tau_{\rm sct} \delta$.

\section{Discussion.}
\begin{figure}[t]
  \includegraphics[width=\columnwidth]{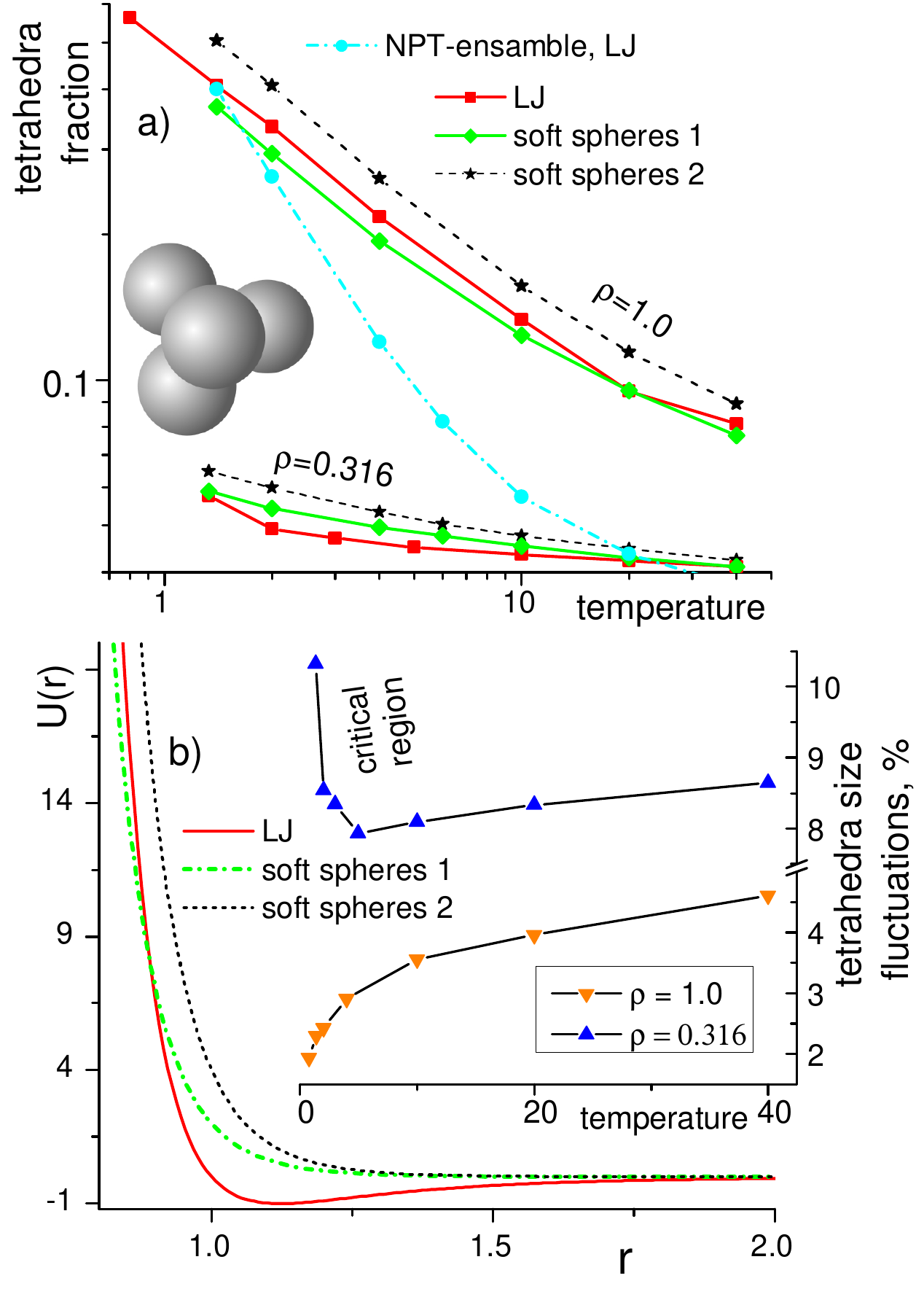}\\
  \caption{(Color online) (a) Tetrahedral fraction vs. temperature for $\rho=1$ (upper three curves) and for (LJ-critical density) $\rho=0.316$ (lower three curves). Particles interact via LJ and soft sphere potentials obtained from LJ by removing the attractive part, type 2, and adjusting the core in addition, type 1 (see Fig.~\ref{fig4}b). The tetrahedral fraction in the NPT-ensemble (LJ-potential) is shown by the dash-dot curve. E.g., at $T=1.362$, the pressure corresponds to the density $\rho=1$; the intersection of the corresponding NVT and NPT concentration-curves in Fig.~\ref{fig4}a at that point demonstrates the stability of the local order with the respect to the ensemble choice far from the LJ-critical point with strong density fluctuations. The same applies for the second (``low density'') intersection in the lower right corner of Fig.~\ref{fig4}a. (b) Main frame: interparticle potentials. Inset: the tetrahedra size fluctuations; they strongly increase in the critical region.}\label{fig4}
\end{figure}
\subsection{Comparison with soft spheres model}

\begin{figure}[t]
  \centering
  \includegraphics[width=0.99\columnwidth]{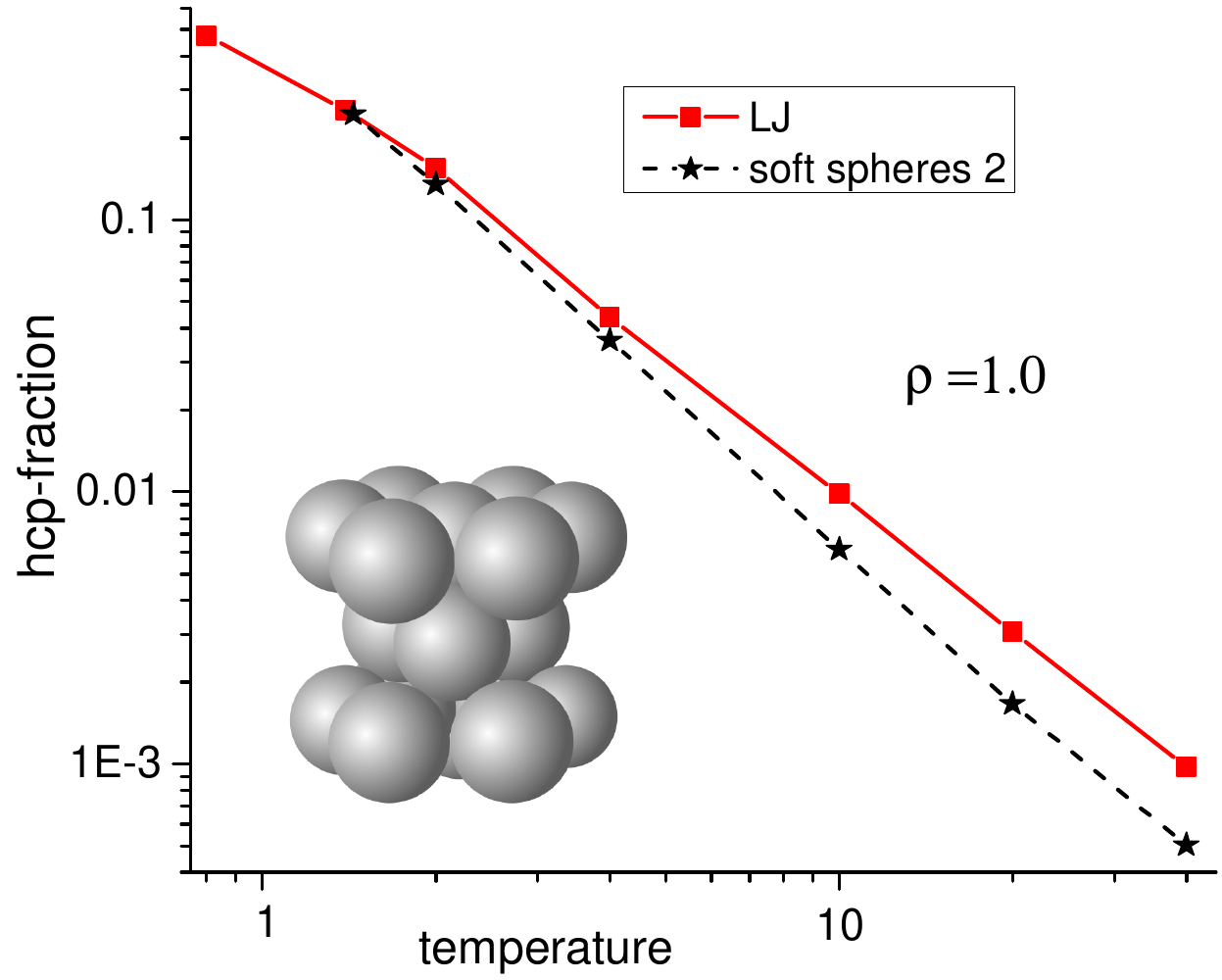}
  \caption{(Color online)Hcp-fraction at density $\rho=1$ in LJ and soft spheres, type 2 (see soft spheres potentials in Fig.~12b).}\label{fig_hcpVStet}
\end{figure}

To test our main results we use the soft sphere model. First we concentrate on parameter domain far from the critical region of the LJ-particle system (soft spheres do not have the critical point). We show here that both hcp and tetrahedra local structure obtained in the frames of LJ model is qualitatively the same as in the soft sphere model, see Fig.~\ref{fig4}. At the critical density we do not observe hcp in both models.. We also check that NVT and NPT ensembles produce similar results for the local cluster concentration in the density region $0.316\leq\rho\leq1$, see, e.g., Fig.~\ref{fig4}a. We also calculated the dispersion of the tetrahedra concentration (smaller or equal to the point-size in Fig.~\ref{fig4}a) in both ensembles and found that even at $\rho=0.316$ it is by the order of magnitude smaller than the average concentration. Similar results we have got for hcp-fraction, see  Fig.~\ref{fig_hcpVStet}. We see in Fig.~\ref{fig_hcpVStet} that the fraction of hcp-clusters at high temperatures in the soft sphere model deviates from LJ stronger than that for tetrahedra in Fig.~\ref{fig_hcpVStet}b. This is so may be because the attractive part of the potential is more important for hcp formation than for tetrahedra.

Going to the critical region in the LJ-system we still observe no hcp-clusters but see the tetrahedra with fluctuating sizes (inset in Fig.~\ref{fig4}b where again the dispersion of the tetrahedra fraction is smaller or equal to the point-size); the results obtained there within NPT and NVT ensembles give different tetrahedra concentration but of the same order. For example, $T=1.33$ and $\rho=0.316$ corresponds $P=0.14$ and the tetrahedra fraction in NPT and NVT are equal to 6.4$\%$ and 5.8$\%$ correspondingly.  We leave the overwhelming examination of the local structure closer to the critical point where the fluctuations exceed 15$\%$ for the forthcoming paper.

It follows that strong tetrahedral order stability is not just model-dependent effect but rather universal feature of simple liquids related mostly to the repulsive shoulder of the intermolecular potential.

\subsection{Virial expansion and tetrahedral contribution to fluid thermodynamics}

\begin{figure}[t]
  \centering
  \includegraphics[width=0.99\columnwidth]{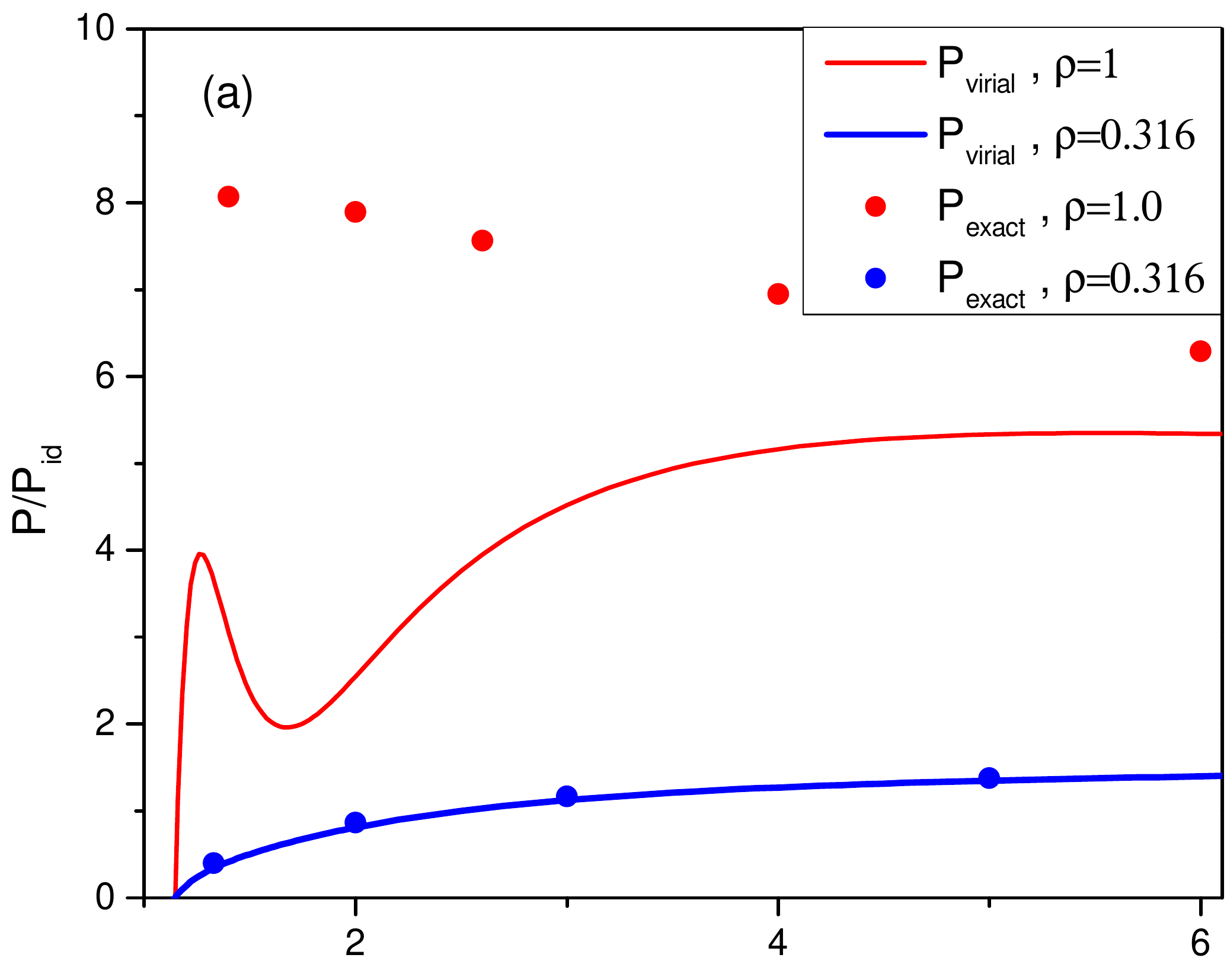} \\ \includegraphics[width=0.99\columnwidth]{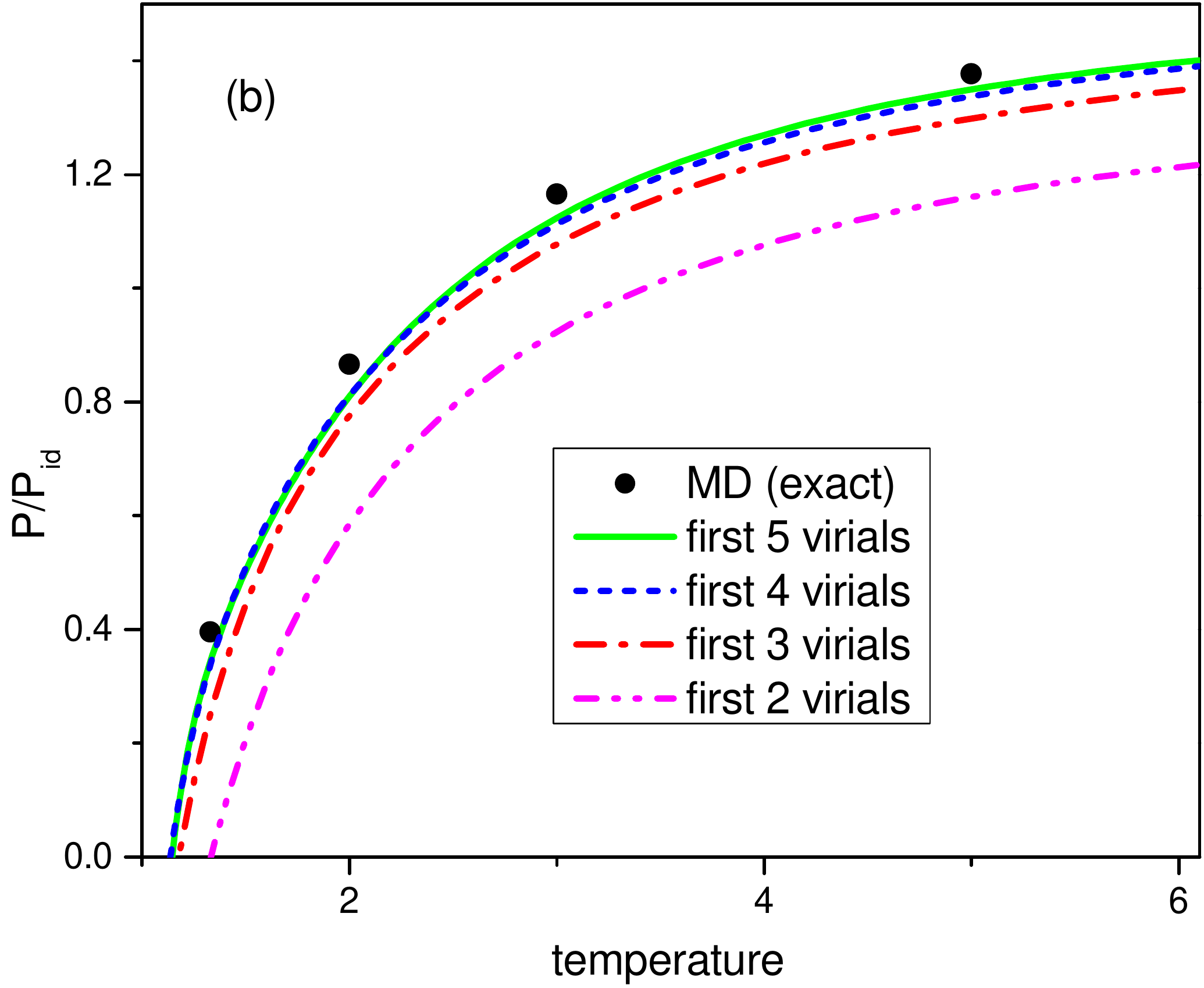}
    \caption{(Color online) (a) The equations of state for Lennard-Jones system calculated by both the exact equation (\ref{equation of state}) and virial expansion (up to $B_5$); (b) The comparison of exact equation of state with first $n=2,3,4,5$ terms of virial expansion at critical density $\rho=0.316.$} \label{fig:virial}.
\end{figure}
Let us discuss the contribution of tetrahedral clusters to fluid thermodynamic. We can try to uncover this problem investigating the virial expansion of the equation of state for system with pair potential $U(r)$ (see, for example \cite{Hill,Hansen}):
\begin{equation}\label{virial_gen}
\frac{P}{{P_{\rm id} }} = \frac{{\beta P}}{\rho } = 1 + \sum\limits_{i = 2}^\infty  {B_i (T)\rho ^{i - 1} },
\end{equation}
where is $\beta=1/kT$; $P_{\rm id}=\rho/\beta$ is the ideal gas pressure. The virial coefficients $B_i(T)$ are defined as $B_1=1$,
$$
B_{i + 1}  =  - \frac{i}{{i + 1}}\beta _i ,\quad i \ge 1,
$$
where the coefficients $\beta_i$ are are the irreducible ``cluster integrals''~\cite{Hill}:
\begin{eqnarray}
 &&B_2  =  - \frac{1}{2}\beta _1  =  - \frac{1}{2}\int {f(r)d{\bf r}} \nonumber \\
 &&B_3  =  - \frac{2}{3}\beta _2  =  - \frac{2}{3}\int {f(r)} f(r')f(\left| {{\bf r} - {\bf r'}} \right|)d{\bf r}d{\bf r'} \\
 && \ldots \nonumber
 \end{eqnarray}
Here $f(r)=\exp(-\beta U(r))-1$ is the so called Mayer function.  The Mayer function (like the pair potential) is far from zero only when particles are situated nearby. So local clustering of particles favors the Mayer cluster integrals. More precisely, the integrands in $B_i$ are noticeably different from zero only if $i$ atoms are close to each other. So $B_i$ represent the contributions of $i$-particle collisions in system thermodynamics.

In order to examine the role of many-particle interactions, we compare the virial expansion (\ref{virial_gen}) with exact equation of state that can be calculated for system with pair potential $U(r)$ as:
\begin{equation}\label{equation of state}
\frac{{\beta P}}{\rho } = 1 - \frac{2}{3}\pi \beta \rho \int\limits_0^\infty  {r^3 g(r)\frac{{dU(r)}}{{dr}}} dr,
\end{equation}
where $g(r)$ is the radial distribution function. For virial expansion we use first five $B_i$ coefficients calculated in~\cite{Dyer}.

Direct calculation shows (see Fig.~\ref{fig:virial}) that contribution of many-particle interactions is essential for supercritical fluids at even low densities and high temperatures where we see local tetrahedra. At low densities, there virial expansion converges to exact state equation very well, we can even extract the four-particle virial $B_4$ defining the contribution from single tetrahedra. We see (Fig.~\ref{fig:virial}b) that this contribution is noticeable, though small enough, at even high temperatures. It is obvious that tetrahedral contribution is even more essential at high densities there finite virial expansion fails and so infinite series of Mayer diagrams, including multi-tetrahedral configurations, must be taken into account.

So, with regard to our results concerning local structure, when we have high enough concentration of local tetrahedra we can conclude that they play significant role in the virial coefficients. We believe that this qualitative conclusion can help our understanding of fluid physics. And we will investigate this issue in more details in the forthcoming paper as the matter of separate work.

\subsection{Trimers, dimers \textit...}

At low density (high temperature) the tetrahedral clusters destroy mostly, but two dimensional and one dimensional symmetric clusters, trimers and dimers, may still survive. In this sense, we can in principle introduce other stages of local order destroy when three dimensional local symmetry changes to two dimensional one, and so on:  hcp-tetrahedra-trimers-dimers-atoms instead of the sequence hcp-tetrahedra-atoms we used. We should emphasize that according to the angular distribution plots, Fig.~\ref{fig:angle}, perfect triangles should destroy more or less at the same temperature (density) as tetrahedra do. Moreover, it seems to us that for 3D one-particle system with isotropic potential the three dimensional local order (tetrahedron) is the matter of prime concern.

\section{Conclusions}
\textit{In conclusion}, we uncover nature of simple supercritical fluid on nanoscales. We show that temperature increase (or/and density decrease) cause two-stage transformation of local order in fluid. On the first stage, at temperatures several times higher the melting temperature the solid-like order disappears. But the fragments of the close packed clusters -- tetrahedra -- survive at temperatures up to the several orders of magnitude higher than the melting temperature. We show that the local density fluctuations in the rare fluid effectively preserve the local symmetry but induce the local cluster size fluctuations. The second stage of the local order transformation corresponds to the total disappearing of tetrahedral clusters and takes place only in the ideal gas limit. We show that the superstability of the tetrahedra clusters in the fluid is not just model-dependent effect but rather universal feature.

\acknowledgments
The work was supported by Russian Foundation for Basic Research (grants  12-03-00757-a, 13-02-91177, 13-02-00579, 13-02-00407 and 11-02-00-341-a), the Grant of President of Russian Federation for support of Leading Scientific Schools No.~6170.2012.2, NSF Grant DMR 1158666, Ural Division of Russian Academy of Sciences (grant RCP-13-P15) and Presidium of Russian Academy of Sciences (program ¹ 12-P-3-1013). We are grateful to Joint Supercomputer Center of the Russian Academy of Sciences and Ural Branch of Russian Academy of Sciences for the access to JSCC and ``Uran'' clusters.


\begin{thebibliography}{99}
\bibitem{note} At the ``dynamical line'' thermodynamic derivatives do not reach a maximum or a minimum as they do at Widom lines.
\end{thebibliography}
\end{document}